\documentclass[a4paper,3p]{elsarticle}
\journal{Computer Physics Communications}
\usepackage{lineno}
\modulolinenumbers[5]
\usepackage{hyperref}
\usepackage{graphicx,caption}
\usepackage{cmap}
\usepackage[T1]{fontenc}
\usepackage[utf8]{inputenc}
\usepackage[english]{babel}
\usepackage{textcomp}
\usepackage{xcolor}
\usepackage{amsmath}
\usepackage{amssymb}
\usepackage{bm}
\usepackage[normalem]{ulem}
\usepackage{physics}
\usepackage{multirow}
\usepackage{hhline}
\usepackage{subfigure}
\usepackage{comment}
\usepackage{float}
\usepackage{multicol}
\usepackage{blindtext}
\usepackage{amsfonts}
\usepackage{empheq}
\counterwithin{equation}{section}
\graphicspath{images}
\biboptions{sort&compress}

\newcommand{\MDcraft}{\texttt{MDcraft}~}
\newcommand{\VD}{VD$^3$~}
\newcommand{\un}[1]{\;\mathrm{#1}}
\begin{document}

\begin{frontmatter}
\title{\MDcraft --- a modern molecular dynamics simulation package with machine learning potentials support}
\author{I.~S.~Galtsov$^{1,2,3}$}
\ead{galtsov.is@phystech.edu}
\author{R.~V.~Muratov$^{2}$} 
\author{G.~V.~Vyskvarko$^{1,2,3}$}
\author{S.~A.~Murzov$^{1,2}$}
\author{S.~A.~Dyachkov$^{1,2}$}
\author{P.~R.~Levashov$^{1,3}$}
\address{$^{1)}$ Joint Institute for High Temperatures of RAS, 13 bldg. 2 Izhorskaya st., Moscow, 125412, Russia}
\address{$^{2)}$ Dukhov  Research Institute of Automatics, 22 Sushchevskaya St., Moscow 127030, Russia}
\address{$^{3)}$ Moscow Institute of Physics and Technology, 9 Institutskii per., Dolgoprudny, Moscow Region, 141700, Russia}

\begin{abstract}{}\noindent
Molecular dynamics is widely used to study various phenomena, such as diffusion, shock wave propagation, and plasma dynamics. A wide range of software packages supports the expanding scope of molecular dynamics applications. However, the quality of simulations depends on force field approximations, ranging from simple models to direct quantum solutions. Recently, machine learning approaches for constructing accurate interatomic potentials have received significant attention. In MDcraft, we integrate these advances into a scalable, physically accurate framework.

MDcraft is a comprehensive, modern molecular dynamics platform. It offers a high-level Python API with a user-friendly, script-based interface. The core simulation algorithms are implemented in C++ to ensure robustness and computational efficiency. MDcraft is built for high-performance computing on modern clusters and supports dynamic domain decomposition and load balancing via the Message Passing Interface (MPI) for scalable parallelization. Additionally, MDcraft leverages multithreading within nodes through standard C++ parallelism, enabling efficient use of heterogeneous architectures.

We demonstrate the code's capabilities through several examples, including the shock response in aluminum, the shock Hugoniot in argon, and the cold curve of copper.

\noindent \textbf{PROGRAM SUMMARY}

\begin{small}
\noindent
{\em Program Title:} MDcraft \\
{\em CPC Library link to program files:} (to be added by Technical Editor) \\
{\em Developers' repository link:} https://github.com/JLab-MatSci/MDcraft \\
{\em Licensing provisions:} GNU General Public Licence 3.0 \\
{\em Programming language:} C++, Python \\
{\em External routines/libraries:} Eigen library, pybind11 library, MPI library, DeePMD, MLIP4
\\
{\em Nature of problem:}
Molecular dynamics simulations for shock wave propagation, diffusion and other tasks \\
{\em Solution method:}
\MDcraft solves equations of motion for atoms given by Newton's second law with conventional and machine learning interatomic potentials. Integration is performed using the conventional Verlet algorithm. The high-performance design of the core molecular dynamics algorithms allows efficient distributed simulations on modern computing clusters. A convenient Python-based user interface enables a broad range of simulations with using complicated processing pipeline and external libraries.\\

\end{small}

\end{abstract}

\end{frontmatter}

\section*{Introduction}
{}
Atomistic simulations have become indispensable tools for probing material behavior at the fundamental scale, with molecular dynamics (MD)~\cite{mdBOOK2022} and density functional theory (DFT)~\cite{martin2020electronic} serving as cornerstones of modern computational materials science. MD enables the simulation of time-dependent atomic trajectories by numerically integrating Newton’s equations of motion, providing detailed spatial and temporal resolution. When combined with first-principles electronic structure methods, such simulations offer high accuracy in predicting energies and interatomic forces without relying on empirical assumptions. However, the computational cost of DFT scales cubically (or worse) with system size, severely limiting its applicability to small-scale systems and short simulation times. In contrast, classical MD employs empirical interatomic potentials (IAPs) or force fields—such as EAM~\cite{Becquart01061997} or Tersoff~\cite{NGUYEN2017113}—that allow for large-scale and long-timescale simulations, but often lack sufficient accuracy and transferability across diverse chemical environments.

In recent decades the number of applications of classical molecular dynamics (MD) method has grown dramatically with the exponential rise of computational capabilities. The basic MD method for short-range interactions scales linearly as $O(N)$ in the number of atoms $N$ and may be efficiently parallelized. As a result, the limitations of the method in time and space have been extended to microseconds and micrometers (continuum scales) with simulations of billions of atoms. At these scales it is possible to simulate natural phenomena with atomistic resolution, which enables us to study plasticity in metals~\cite{bulatov}, chemical phenomena~\cite{GISSINGER2017211}, or biological molecules, membranes and cells~\cite{membranes}.

In materials science, MD is extensively used to investigate mechanical, thermal, and transport properties such as elasticity, thermal conductivity and diffusion coefficient \cite{DiffusionReview2021}. With millions of atoms in a computational domain, MD approach allows for phase transitions research  under varying temperature and pressure conditions~\cite{BPNN_2007,IceMelt2024}. It enables the study of defects, grain boundaries, and nanostructured materials, providing insights into deformation mechanisms and failure processes. MD is also an instrument for simulating cluster systems~\cite{Poletaev2024}, where finite atomic or molecular aggregates exhibit size-dependent properties relevant to catalysis and nanotechnology. In biophysics, MD simulations now help improving molecular docking and drug design~\cite{Hospital12122015} and allow exploring details of protein-protein interactions~\cite{wu2022application}. Additionally, MD is employed in extreme states of matter physics to simulate shock waves (SW)~\cite{Zhakhovsky:PRL:2011}, capturing the propagation of stress and energy through materials on ultrafast timescales. In plasma physics \cite{Cimarron_plasmas_2012,sarkas2022}, it contributes to modeling dense plasmas and ion behavior in partially ionized states, complementing continuum approaches.

The growing number of applications is supported by a plenty of MD software packages. One may notice, that biological and chemical applications have the largest demand that is satisfied by \texttt{AMBER}~\cite{amber2012,amber2023}, \texttt{CHARMM}~\cite{charmm2024}, \texttt{Desmond}~\cite{desmond2006} (as a part of Shr\"{o}dinger package), \texttt{GENESIS}~\cite{genesis2024}, \texttt{GROMACS}~\cite{gromacs2015} and \texttt{NAMD}~\cite{namd2005} packages. However, simulation of complex molecules require specific force fields or even quantum mechanical calculations: the development of such tools requires years of study, and is also a reason for commercial usage of some packages. In materials science and chemistry \texttt{LAMMPS}~\cite{lammps2022} is used as conventional software package allowing to model atomic systems at largest scales and to perform various analysis.

In recent years, machine learning (ML) methods for evaluating interatomic forces~\cite{BPNN_2011,MAISE_2021,GAP2010,DeePMD2018,WANG2024109673,ham2025} have attained growing attention. As long as traditional approaches face a persistent trade-off between computational efficiency and predictive power, they are often too simplistic to capture complex bonding environments, whereas \textit{ab initio} methods, though accurate, are prohibitively expensive for extensive sampling. ML has emerged as a strong paradigm, capable of learning complex mappings between atomic configurations and energetic properties from high-fidelity reference data. Nowadays, it is possible to use ML-based collective force field evaluation algorithms instead of conventional pairwise ones so that modeling of atomic systems with precision close to quantum mechanical becomes possible. The computational cost of such potentials may be several orders larger, but still it is possible to simulate dynamics of millions of atoms~\cite{millionsAtoms} instead of thousands with DFT approaches, because of preserving the linear scaling in the number of atoms. Some of the aforementioned packages already support machine learning potentials.

At the core of all machine learning interatomic potentials (MLIP) or machine learning molecular dynamics (MLMD) approaches lies the choice of descriptors encoding the local atomic environment. These descriptors transform geometric information about neighboring atoms into a numerical representation suitable as input for a trainable model. High-quality descriptors must satisfy fundamental physical symmetries---namely invariance under global translations and rotations, as well as permutation of chemically equivalent atoms. Among the most widely used descriptors are atom-centered symmetry functions~\cite{BPNN_2011}, smooth overlap of atomic positions (SOAP)~\cite{SOAP}, spectral neighbor analysis method (SNAP)~\cite{SNAP} and moment tensor potentials (MTP)~\cite{MTP,Novikov_2021}.

Recently, there has been growing interest in equivariant MLMD frameworks~\cite{En2024,En2025}. Unlike conventional approaches that rely solely on scalar (invariant) descriptors, equivariant models explicitly respect the geometric nature of atomic interactions by processing vectors, tensors, and higher-order features in a rotationally equivariant manner. This enables a more faithful representation of directional bonding, anisotropic forces, and many-body effects, leading to significantly improved accuracy and transferability of the resulting potentials.

Beyond the choice of descriptors, MLMD approaches also differ in the type of a regression model used to construct the interatomic potential. A comprehensive analysis and benchmarking of both descriptors and model architectures can be found elsewhere (see~\cite{WANG2024109673,ham2025} and references therein). Here, we briefly mention only a few widely adopted model types: artificial neural networks (ANN)~\cite{BPNN_2011,MAISE_2021,DeePMD2018,unke2021spookynet,ARTRITH2016135,En2024}, polynomials~\cite{MTP,Novikov_2021,Allen_2021} and Gaussian regression~\cite{GAP2010}.

The pipeline of modern machine learning tools is based on Python scripting, as soon as it allows a convenient organization of high performance modules in a single program. It provides a reasonable flexibility in adjusting the learning algorithm that may be run on modern hardware without low-level coding. The aforementioned MD packages have been developing for decades and well fit the most user demands. However, complicated functionality may require a lot of time for users to learn the API to include some specific tools. It is worth noting the Atomic Simulation Environment (ASE) \cite{ASE} as a well-established Python platform for addressing a wide range of atomistic modeling tasks, including molecular dynamics simulations. Several test examples in \MDcraft leverage the convenient functionality as given by ASE.

\MDcraft package is developed to provide the user with high-level Python API for MD simulations. It allows to use any Python library during MD simulation for analysis, post-processing or data-mining. At the same time, it allows to run MD algorithms in parallel on modern clusters with dynamic domain decomposition and load balancing. It supports both classical pair-style potentials and modern ML potentials. The available flexibility allows to write a Python script for complicated simulations like moving window (MW)~\cite{MURZOV:2024}, combined DFT and MD, or using various potenials without modifying the core modules. Some usage examples are provided.

\section{MDcraft package description}
\label{sec:MDcraft}

The software package can be represented as a hierarchy of loosely coupled modules, each designed to address a specific set of thematic tasks. The methods and algorithms which are essential for  the MD simulation are separated into the following core modules:   
\begin{itemize}
    \item \texttt{data} --- Module for atomic data representation.  
    \item \texttt{decomposition} --- Algorithms for inter-node communications and parallel distributed task execution across multiple compute nodes with load balancing.  
    \item \texttt{io} --- Module for data input/output (including parallel operation) in various formats.  
    \item \texttt{lattice} --- Module for defining the computational domain and atoms positions.  
    \item \texttt{neibs} --- Module for the neighbor list construction.  
    \item \texttt{solver} --- Module for integrating classical MD equations of motion.  
    \item \texttt{tools} --- Module for auxiliary utilities and tools.
\end{itemize}

Each module may, in turn, consist of submodules and components. A component represents a class of objects solving a single specific task. Several thematically related components may be grouped into a module. Furthermore, components may utilize other components to accomplish their tasks. Thus, the greater the number of components from one module used in implementing another module, the stronger the coupling between them.   

The proposed modules ensure relatively weak coupling, allowing the functionality of many modules to be used independently. For example, lattice generation is unrelated to simulation methodology, enabling a lattice to be constructed and visualized prior to the modeling phase. Domain decomposition across compute nodes is likewise decoupled from simulation methods, but provides a data access interface for arrays requiring processing during simulation. Nevertheless, the proposed architecture may present hidden challenges to be addressed during the implementation.   

All modules feature a corresponding Python interface implemented via the \texttt{pybind11}~\cite{pybind} library, enabling the use of third-party libraries during simulations. For instance, this interface allows: application of standard mathematical algorithms at various design and simulation stages using \texttt{SciPy}~\cite{scipy}, visualization via \texttt{matplotlib}~\cite{matplotlib}, integration of machine learning algorithms using libraries such as \texttt{scikit-learn}~\cite{sklearn}, \texttt{pytorch}~\cite{pytorch}, \texttt{tensorflow}~\cite{tensorflow}, and others.

\subsection{Data structures}

\begin{figure}[t]
\centering
\includegraphics[width=0.99\textwidth]{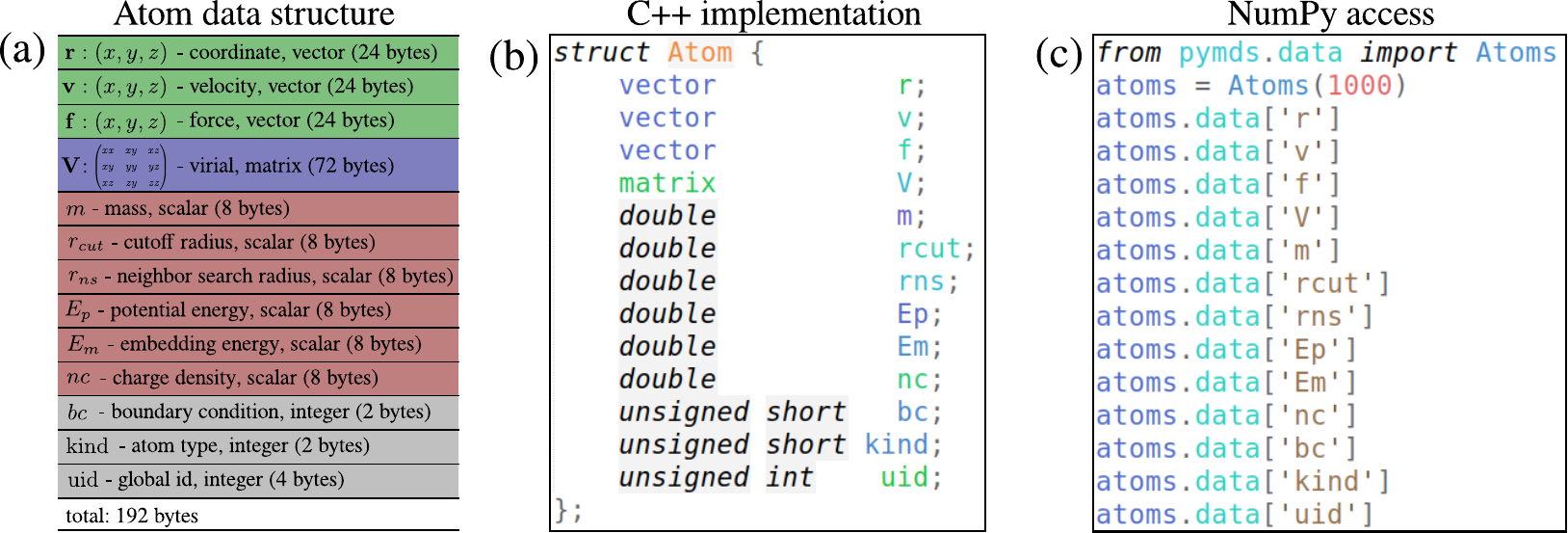}
\caption{(a) Basic Atom data structure suitable for a large number of applications. (b) C++ implementation of the Atom data structure. (c) Access fields of the Atom data structure using Python. }
\label{fig.atom-structure}
\end{figure}

The \texttt{Atom} data type for storing atomic information is defined in the \texttt{data} module. It is implemented as a \textit{Plain Old Data (POD)} structure containing a minimal set of fields required for a complete description of an atom's state, as illustrated in Figure~\ref{fig.atom-structure}(a,b). Atoms are stored in a corresponding C++ \texttt{std::vector} container, ensuring all atom-processing algorithms are optimized for this specific data type.

The atom structure includes both the \textit{physical fields} (position, velocity, force, mass, etc.) and \texttt{technical fields} (neighbor search radius, atom type, etc.). Vector quantities are processed using the \texttt{Eigen} library~\cite{Eigen} that provides vector arithmetic operations, scalar-vector multiplication, matrix-vector multiplication, and other useful algorithms for vector processing. 

A critical feature of the \MDcraft software package is the direct memory access to the atom array through its Python interface: the \texttt{.data} operator converts the atoms data to a NumPy array by constructing proper metadata to the underlying raw bytes array. This enables direct read/write operations to the memory space of the \texttt{std::vector} container. Thus, one may process data using both the native MDcraft algorithms and third-party libraries via Python interoperability.

\subsection{Hybrid parallel algorithms}

Presently, multi-core processors and computing accelerators are employed in personal computers and supercomputers around the world. In the context of addressing intricate computer simulation problems, the utilization of parallel computing should not be regarded as a mere option but rather as an imperative.

\begin{figure}[t]
\centering
\includegraphics[width=0.99\textwidth]{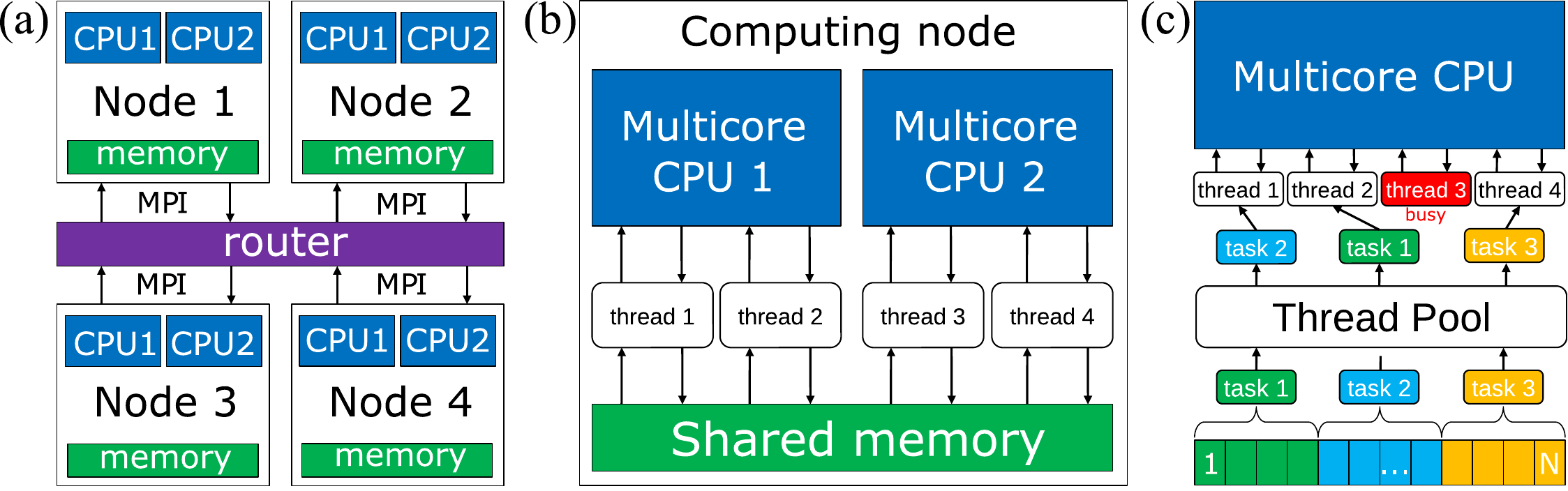}
\caption{(a) A distributed memory system scheme consisting of 4 computing nodes communicating via MPI. (b) A shared memory system scheme within a computing node with 2 multicore CPUs. (c) The process manager (thread pool) usage scheme for handling array elements in parallel. }
\label{fig.computer-architecture}
\end{figure}

With respect to the management of memory, computers can be categorized as distributed memory systems and shared memory systems. Distributed memory systems are characterized by the distribution of memory across multiple components or processes within a computer system (see Fig.~\ref{fig.computer-architecture}a). In contrast, shared memory systems are designed to allocate memory to specific processes or tasks within the system, enabling efficient utilization of resources and facilitating concurrent execution of operations (see Fig.~\ref{fig.computer-architecture}b). Memory is directly accessible to any processor in the case of shared memory and inaccessible in the case of distributed memory. In essence, a system with distributed memory is a cluster whose nodes are interconnected by a network. Personal computers are typically shared memory systems. Shared memory can be physically distributed; however, for processors, it must represent a single address space. The specific type of memory operation defines the chosen technique for program parallelization.

When working with distributed memory, the MPI API (Message Passing Interface Application Programming Interface) is traditionally used when a cluster node needs to access memory areas that are not directly accessible to it. MPI facilitates data transfer between nodes in the form of messages (see Fig.~\ref{fig.computer-architecture}a). Within a node, multithreaded programming technologies, such as OpenMP and Intel TBB, are commonly used (see Fig.~\ref{fig.computer-architecture}b). The \MDcraft package implements both approaches, which can be used simultaneously in a hybrid parallelization regime: MPI works between the nodes of a computing machine, while thread parallelization using Intel TBB works inside the node.

Let us consider parallelization within a single node of a cluster or for a system with shared memory. As mentioned above, the program uses a process manager based on some implementation (see Fig.~\ref{fig.computer-architecture}c). The process manager (thread pool) allocates a user-specified number of processes and assigns incoming tasks to them as they become available. A task may involve processing $m$ elements of an array, for example. A list of $K$ tasks can be created by dividing the array of $N$ elements into $K$ parts. If the number of tasks exceeds the number of processes handled by a manager, the additional tasks will wait in a queue. The number of processes is typically set as a multiple of the number of processor cores. This approach balances the load when working with shared memory. Since different parts of the array may be processed at different speeds, dividing the array into a number of parts equal to the number of processes may result in one process taking longer than the others. When the number of tasks exceeds the number of processes, the processes receive random parts of the array, distributing the load more evenly. At the same time, if the amount of work within a task is too small compared to the task's start-up time, the efficiency of parallelization decreases sharply. Therefore, the number of tasks for parallel processing should be chosen carefully to take into account both mentioned aspects.

\subsection{Dynamic domain decomposition with load balancing}

In distributed memory systems (clusters, see Fig.~\ref{fig.computer-architecture}a), data is physically separated, i.e., decomposed. In our case, the data element is represented by it's fields of the atom structure shown in Fig.~\ref{fig.atom-structure}a. An increase in the number of cluster nodes provides more memory for the storage, allowing for a greater number of atoms in the calculation. The data of a given atom is accessible directly only to one node of a parallel machine. Nodes process data at different speeds depending on the mathematical model (e.g., the type of interatomic potential) and the computing node's characteristics. When computational work is unevenly distributed, the calculation efficiency inevitably drops. It happens because lightly loaded nodes are idle while heavily loaded ones are busy. The delays are caused by the need to access updated data elements in other nodes of the cluster, i.e., at a synchronization point. Obtaining data about atoms incurs additional time costs due to data transfer over the network connecting the nodes of the parallel machine. To minimize these costs, it is advisable to ensure that the decomposition algorithm distributes the computational work evenly among the nodes with minimal data transfer between them. Such a decomposition algorithm ensures efficient use of resources and time for problem solving. Therefore, the data decomposition algorithm is an important part of any program performing a simulation using massively parallel machines. With optimal data partitioning and the well-optimized calculation algorithm within each node, calculation time is reduced proportionally to the number of processors involved.

A number of studies have proposed principles for constructing the most efficient data redistribution methodology to achieve the best calculation performance. The decomposition update procedure must be aligned with the load balance data. It is claimed in Refs.~\cite{cybenko1989dynamic, willebeek1993strategies, hu1998optimal} that the optimal decomposition balancing algorithm should be of the diffusion type: the computational load should be redistributed similarly to heat in the process of heat conduction. The useful load is measured either proportionally to the number of elements in the computational subdomains~\cite{ferrari2009new, oger2016distributed}, or, more naturally, by the time spent on calculations~\cite{sbalzarini2006ppm, fleissner2008parallel, zhang2009fast, zhakhovskii2005new, fattebert2012dynamic}. 

The algorithm used in the software package operates on the basis of dynamic decomposition according to the Voronoi diagram~\cite{koradi2000point, zhakhovskii2005new,fattebert2012dynamic, fu2017physics, Egorova:CPC:2019, Muratov:CPC:2023}. In accordance with~\cite{cybenko1989dynamic, willebeek1993strategies, hu1998optimal}, the load is measured as the ratio between the ``useful'' calculation time and the total calculation time per step. A specific load balancing algorithm \VD (Voronoi Dynamic Domain Decomposition) is selected, originally proposed for the molecular dynamics method~\cite{zhakhovskii2005new}, and later extended by us to the smoothed--particle hydrodynamics (SPH) method~\cite{Egorova:CPC:2019} and to mesh hydrodynamic methods~\cite{Muratov:CPC:2023}.

\begin{figure}
\centering
\begin{minipage}{0.6\linewidth}
\centering
\includegraphics[width=\linewidth]{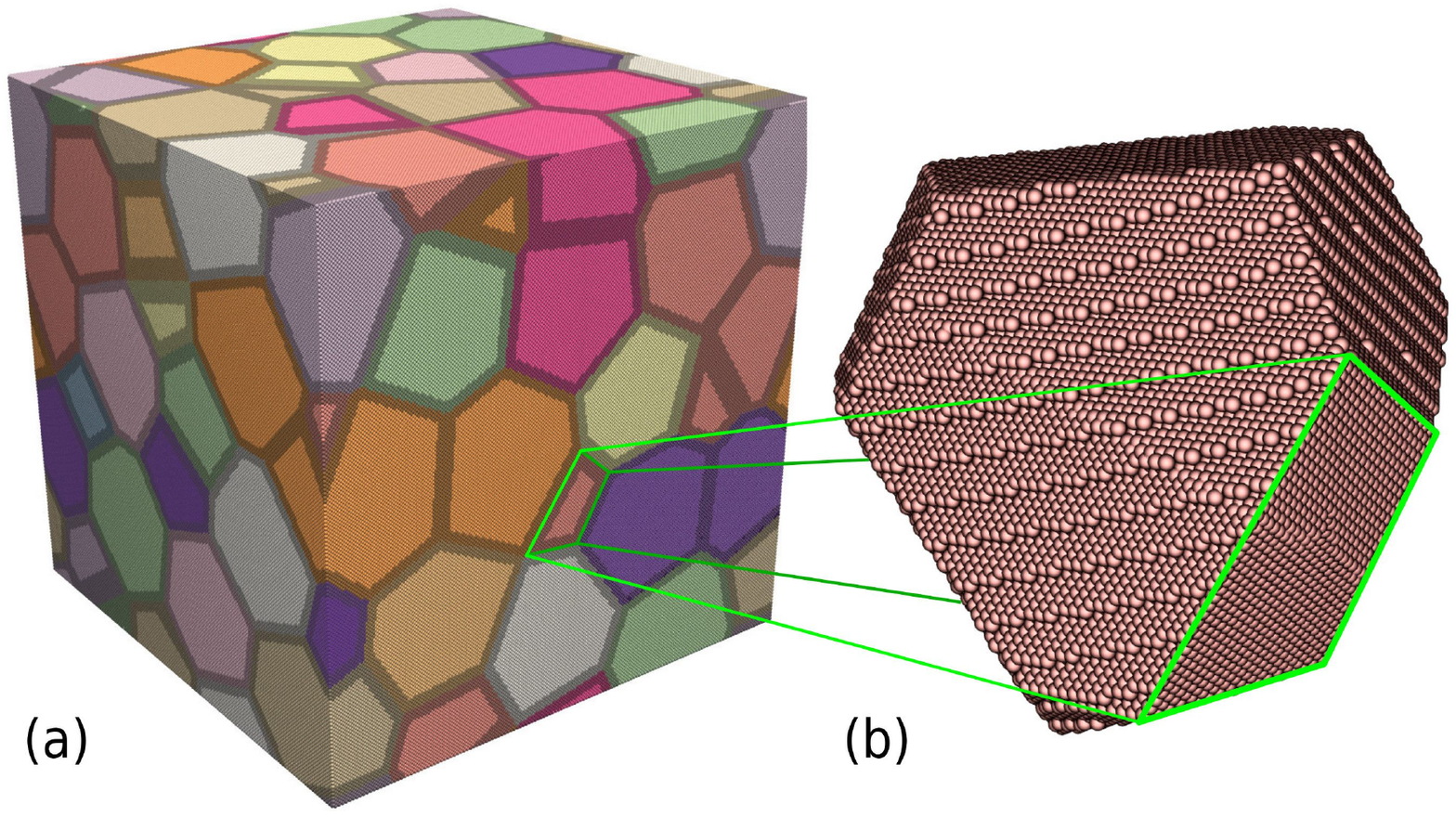}
\caption{(a) Decomposition using a Voronoi diagram with data exchange layers. (b) Voronoi domain containing atoms handled by a single MPI process. \label{fig:VD3}}
\end{minipage}
\end{figure}

According to~\cite{du1999centroidal}, the Voronoi diagram is a partition of a closed subspace~$\bar\Omega\in\mathbb{R}^n$ into~$N_{\hat V} $  parts~$\{V_k\}_{k=1}^{N_{\hat V}}$ (cells) based on the distance to a certain set of points~$\{G_k\}_{k=1}^{N_{\hat V}}$, called the centers of the Voronoi diagram:
\begin{equation}\label{eq:VoronoiDefinition}
 \hat V_k = \{\left. \vec{r}\in\Omega\right.\,:\,|\vec{r} - \vec{g} _k | <| \vec{r} - \vec{g} _l |,
 \,l = 1,\, \ldots,\, N_{\hat V}, l \ne k \},
\end{equation}
where~$\vec{g}_k$ is the radius vector of the point~$G_k$. In other words, in definition~\eqref{eq:VoronoiDefinition}, it is stated that each cell of the Voronoi diagram is a convex polyhedron that contains all points in space whose distance from the center of this cell is less than the distance from the center of any other cell. 

Suppose that the object of mathematical modeling is a body or a set of bodies represented by a set of atoms. Suppose that~$N_p$ computational nodes are available. Let us distribute the set of points ~$\{G_k\}_{k = 1}^{N_p}$ in the region occupied by the bodies. Establish a one-to-one correspondence between the nodes $k = 1, 2, \ldots, N_p$ and the elements of the set $\{G_k\}_{k = 1}^{N_p}$. The points~$\{G_k\}_{k = 1}^{N_p}$ uniquely determine the corresponding Voronoi diagram~$\{V_k\}_{k=1}^{N_{\hat V}}$. We will place the data associated with the elements in the cell of the diagram~$\hat V_k$ in the memory of process~$k$. If an element is located on the boundary between $\hat V_k$ and $\hat V_l$, we consider it to belong to the subregion $\hat V_{\min\{k,l\}}$. Figure~\ref{fig:VD3} illustrates domain decomposition using a Voronoi diagram. The atoms at the boundaries of the Voronoi cells, which participate in exchange operations between MPI processes, are highlighted in dark color.
The calculated elements that belong to a specific cell are stored in memory and updated during simulation by the process corresponding to that cell. Since the set of centers uniquely defines the cells, the cells can be shifted by moving the centers. Shifts in \VD occur both with the movement of atoms within a cell and under the balancing shifts of an algorithm that equalizes the load between cells~\cite{zhakhovskii2005new, Egorova:CPC:2019, Muratov:CPC:2023}.

\subsection{Neighbor lists construction}

Molecular dynamics simulations with short-range interaction potentials require identifying atoms within the interaction cutoff radius (``neighboring'' atoms). A brute-force search for neighbors $j$ of a given atom $i$ among $N$ atoms requires $\mathcal{O}(N)$ operations. Identifying all interacting pairs can require up to $\mathcal{O}(N^2)$ operations.  

This computational cost can be reduced to $\mathcal{O}(N)$ by using a precomputed \textit{Verlet list} (neighbor list) of candidate atoms within an extended search radius slightly larger than the atomic size, where the list length is significantly smaller than $N$~\cite{allen1989computer}. In practice, a typical Verlet list contains about 100 neighbors per atom.  

The Verlet list may be updated once in several timesteps. While the list is traversed at every timestep to identify interacting pairs (due to particle displacements), the list itself is only rebuilt when particles move beyond the neighbor threshold or domain decomposition is updated.  

For each atom $i$, the Verlet list contains indices of atoms within the neighbor search radius $r_{ns}^i$, defined as:  
\begin{equation}\label{eq:rhrz}
r_{ns}^i = (1 + \beta)r_{cut}^i,
\end{equation}  
where $r_{cut}^i$ is the interaction cutoff radius defined by interatomic potential, $\beta > 0$ is safety buffer coefficient. This buffer ensures interacting atoms remain in the Verlet list between updates.

\begin{figure}
\centering
\begin{minipage}{0.49\linewidth}
\includegraphics[width=\linewidth]{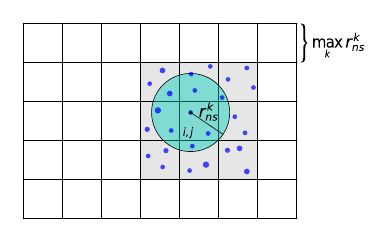}
\caption{Neighbor list construction. Virtual Cartesian grid covers a computational domain.\label{fig:nlist}}
\end{minipage}
\end{figure}

The program employs the following algorithm for neighbor list construction. First, the maximum neighbor search radius $R$ is determined for all particles: $R = \max_k r^{k}_{ns}$. Next, a virtual Cartesian grid is constructed over the computational domain, with cell size equals to $R$ (see Fig.~\ref{fig:nlist}). This grid fully covers the simulation region. For each atom, we can explicitly determine a three‑dimensional cell index corresponding to its location. In fact, in a single pass through all atoms, they can be distributed among the cells of this virtual grid. The number of cells in the virtual grid is significantly smaller than the number of atoms. This atom‑to‑cell assignment can be regarded as a form of bucket sort, which scales as $O(N)$ and exhibits excellent parallelization potential in a multithreaded environment.

Subsequently, the neighbor search is performed for each atom. If an atom resides in a cell with index$(i, j)$, the neighbor search can be limited to atoms located within a $3 \times 3 $ cell template---that is, in cells with indices $i' \in \{i - 1, i, i + 1\}$, $j' \in \{j - 1, j, j + 1\}$ (see Fig.~\ref{fig:nlist}). This approach holds for the two‑dimensional case; in three dimensions, it requires examining all atoms within a $3 \times 3 \times 3$ block (27 cells) of the virtual grid. Assessing the algorithm's performance, the number of operations is on the order of $O(N \cdot n)$, where $n$ represents the average number of neighbors per atom (a value typically on the order of a few hundred). This performance is substantially superior to that of a pairwise comparison algorithm, which scales as $O(N^2)$.

\subsection{Molecular dynamics}

The molecular dynamics simulation algorithm with various interatomic potentials support is defined in the \texttt{solver} module. In general, it deals with a system of $N$ atoms within the domain $\Omega$. The total energy of the atoms $E = K + U$, where $K$ is the kinetic energy:  
\begin{equation}  
K = \sum_{i = 1}^{N} \frac{1}{2} m_i \dot{\mathbf{r}}_i^2,  
\end{equation}  
$U$ is the potential energy:  
\begin{equation}\label{eq:Energy_U}  
U = U(\mathbf{r}_1, ..., \mathbf{r}_N).
\end{equation}
The equations of motion for the system are given by Newton's second law \cite{Yip:2005}:  
\begin{equation}  
\label{eq:MD-system}  
m_i \frac{d^2 \mathbf{r}_i(t)}{dt^2} = \mathbf{f}_i = -\nabla_{\mathbf{r}_i} U, \quad i = 1, ..., N  
\end{equation}  
where $\mathbf{f}_i$ denotes the net force exerted on atom $i$ by surrounding atoms. 

The system~\eqref{eq:MD-system} comprises $6N$ equations for particle position and velocity components. Integration is performed using the conventional Verlet algorithm, at it is symplectic, preserving phase space volume and exhibiting excellent long-term energy conservation properties. By solving this system the $NVE$ ensemble is simulated. One should note, that the right hand side of Eq.~\ref{eq:MD-system} may be appended with Langevin force to perform calculations in $NVT$ ensemble with thermostat.

There are main types of boundary conditions implemented: free and periodic. In some cases (for example in Moving Window simulations, as described below) one may use special thermostats to model the inflow/outflow conditions or hard walls.

\subsection{Potentials support}

In the molecular dynamics method, the main computational load is calculating the interaction force between atoms, which is determined by the interaction potential $U$. Modern machine learning-based potential libraries usually have an interface optimized for multi-core processors and graphics accelerators to calculate forces. Nevertheless, classical short-range interaction potentials remain relevant because the performance of calculations with these potentials is significantly higher. This allows for the modeling of systems with a number of atoms exceeding $10^{10}$, which is currently unavailable with neural network potentials. Depending on the problem, it is almost always possible to select potential parameters that best reproduce one or more key material properties, such as vacancy formation energy, melting point, surface tension coefficient, sound speed, etc. Thus, developers can adjust the parameters to optimize interatomic potentials for specific classes of problems.

To calculate the forces, the following derivative must be calculated:
\begin{equation}
  {\bf f}_i = -\sum_{j \neq i} \frac{\partial V(r_{ij})}{\partial {\bf r}_i}
            = \sum_{j \neq i} \left.
             \frac{\partial V(r)}{\partial r}\right|_{r=r_{ij}}
             \frac{{\bf r}_{ji}}{r_{ji}}.
\end{equation}

\MDcraft supports the main forms of classical short-range potentials, such as Lennard--Jones potential or EAM potential~\cite{Yip:2005}.

\subsection{Performance}

The performance of the implemented parallel algorithms was evaluated using a standardized molecular dynamics test case. The simulation employs a simple LJ potential with atoms initially arranged in a lattice structure with minor positional perturbations. A cubic domain with periodic boundary conditions in all directions is modeled. The test executes 300 time steps, with neighbor list updates---including complete reconstruction of the Verlet lists and neighbor search---performed every 10 steps. The total system size equals 13 million atoms.

\begin{figure}
\begin{minipage}{0.49\linewidth}
\includegraphics[width=\linewidth]{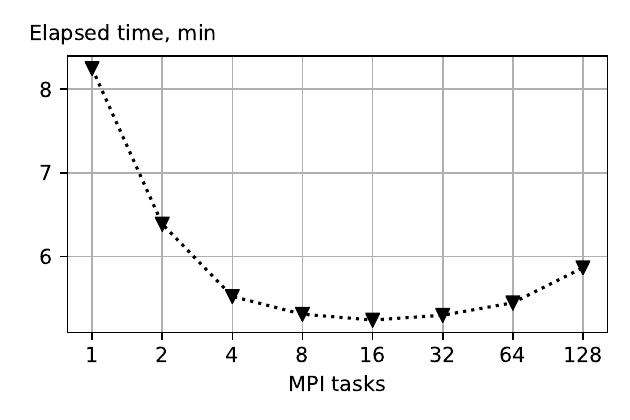}
\end{minipage}
\hfill
\begin{minipage}{0.49\linewidth}
\includegraphics[width=\linewidth]{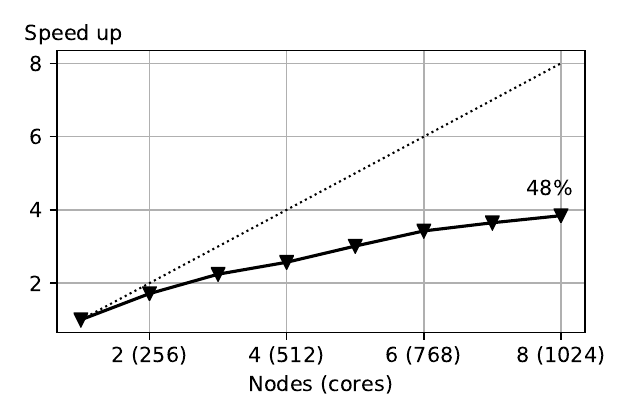}
\end{minipage}
\caption{Performance benchmark results: exectution time on a single node with varying numbers of MPI-tasks (left), and scalability from 1 to 8 nodes (right).\label{fig:performance}}
\end{figure}

Initial benchmark focused on identifying the optimal hybrid parallel configuration within a single computational node. The cluster nodes feature 256 CPU cores, making powers of two natural choices for MPI task counts. The results, presented in Figure \ref{fig:performance}  (left), reveal a non-trivial relationship between MPI processes and threads. The optimal performance was achieved using 8 to 16 MPI tasks with a corresponding number of threads per task. A significant performance improvement was observed when transitioning from a purely multithreaded execution to using 2 MPI processes. This performance jump is attributed to Non-Uniform Memory Access (NUMA) effects. In a purely multithreaded environment, memory access patterns may become suboptimal as the threads spanning multiple NUMA domains compete for remote memory resources. Utilizing 2 MPI processes effectively binds each process to a separate CPU socket, localizing memory access and mitigating NUMA-related contention. Further performance gains with additional MPI tasks are less pronounced and more complex to explain, likely involving a balance between improved memory locality and increased communication overhead.

Subsequent tests evaluated scaling across 1 to 8 compute nodes, employing 8 MPI tasks per node as determined by the single-node optimization. Results are shown in Figure \ref{fig:performance} (right), with dashed lines indicating ideal linear scaling. The implementation achieves approximately 50\% parallel efficiency when utilizing 8 nodes (1024 CPU cores). This level of scalability is considered reasonable for complex molecular dynamics simulations with frequent neighbor list updates and communication patterns.

\section{Molecular dynamics with machine learning potentials}
\label{sec:mdmlp}

The current implementation of \MDcraft supports three types of machine learning interatomic potentials. For the first, a neural network potential based on Behler-Parrinello symmetry functions~\cite{BPNN_2011} and constructed within the \texttt{MAISE}~\cite{MAISE_2021}, we present our custom implementation which uses automatic differentiation for the reconstruction of energies and forces during molecular dynamics simulations. Detailed specifications are provided in the following section~\ref{sec:BPNN}. For the two remaining potential types—MTP~\cite{MTP} and deep learning potentials~\cite{DeePMD2018}—for now we only support the implementation via direct calls to functions from external libraries as given below.

The MTP potentials were developed by Shapeev's group within the open-source \texttt{MLIP} package~\cite{MLIP2021}. The framework of MTP was extensively used for a number of applications~\cite{mlip2020nat, mlip2022nat, mlip2023nat}. This form of potentials is built upon quite intricate tensor convolutions of relative atomic distances.

We also exploit the functionality of \texttt{DeePMD} package for deep potentials support. This type of ML potential has the most complictated atomic descriptor architecture among the three mentioned ones. It supports so called radial and angular embedding matrices \cite{Wen2022} to construct local atomic environment in much detail and consequently describe even rather exotic domains of the system's phase space. This methodology was intensively exploited for a variaty of tasks~\cite{hedman2024dynamics, wieser2024machine}. The described deep-learning potentials were also carefully tested and compared to the aforementioned MTP potentials~\cite{chandran2024comparative}. This study demonstrates that for the same TiAlNb system, the MTP and DeePMD potentials, trained on identical data, exhibit a fundamental trade-off between accuracy and computational efficiency. Consequently, there is no universally superior potential, as their comparative performance dictates the choice of MTP for high-speed simulations versus DeePMD for tasks demanding the highest precision. The computational efficiency of MTP stems from its architecture based on the rapid computation of polynomial basis functions, whereas DeePMD requires sequential computations through multiple layers of a deep neural network, making it substantially more computationally expensive.

The detailed training procedures for both MTP-based descriptors and DeePMD embedding matrices are thoroughly presented in the literature and official documentation. We do not reproduce the formalism of either methods here, as (i) our current implementation provides only a wrapper around the internal functions of these packages, and (ii) the core concepts—namely, the use of atomic descriptors to represent system configurations as input to machine learning potentials, and the subsequent reconstruction of energies and forces from the trained models—are comprehensively illustrated in the following section with an example of a more simplistic architecture of Behler-Parrinello neural networks.

\subsection{Molecular dynamics with Behler-Parrinello neural networks}\label{sec:BPNN}
This section contains the basic equations that are used in Behler-Parrinello neural network (BPNN) approach \cite{BPNN_2007,BPNN_2011}. Our current implementation is made for BPNN MD and uses potentials in the format written by \texttt{MAISE} package \cite{MAISE_2021}. In the BPNN there is a number of symmetry functions $\{\mathfrak{G}\}$ which serve as descriptors for each of $N$ atoms in a system. A separate NN is built for each atom with the input layer size equal to the number of symmetry functions. The main purpose of $\mathfrak{G}$-functions is to convert spatial atomic configurations into a form appropriate for feeding the used NN model. In principle, one could imagine feeding simply $3N$ scalar numbers standing for the full list of atomic coordinates as an input of the large input layer of a single feed-forward NN. Not only this approach inpractical due to the neccesity of retraining the NN for a different system size $N$, but it also lacks atomic permutation invariance. This was the main reason for Behler \cite{BPNN_2011} to present an alternative idea of encapsulating the high-dimensional information of the whole system into a small number of symmetry functions $\{\mathfrak{G}\}$.

Here we mostly follow the original paper \cite{BPNN_2011} with several comments connected with \texttt{MAISE} implementation details. First, the total potential energy of a system is written in the additive form

\begin{equation}\label{eq.E_as_sum}
E = \sum_{i = 1}^{N} E_i,
\end{equation}
where $E_i$ is the potential energy of the $i$-th atom. The success of this energy representation relies on the locality of interactions: only atoms within a certain radius $R_\mathrm{c}$ (hereafter referred to as a cutoff radius) do contribute to the forces acting on a given atom $i$. Traditionally, atoms within the cutoff radius of an atom $i$ are called the neighborhood of atom $i$. We will use $\mathfrak{N}(i)$ notation for the neighborhood of atom $i$. A nice illustrative Figure 2 of the Behler's work \cite{BPNN_2011} shows the architecture of the full high-dimensional NN model. Here we would like to focus on the third column in this Figure, which stands for individual atomic NNs. The input layer of $i$-th atom NN is a vector of several values $\{\mathfrak{G^\mu(i)}\}$ with $\mu = 1, 2, ..., G_\mathrm{i}$ and $G_\mathrm{i}$ equals to the full number of symmetry functions for atom $i$. There are two types of symmetry functions: $R_\mathrm{i}$ radial ones (we denote this subset as $\{\mathfrak{R^\mu}(i)\}$, $\mu = 1, 2, ..., R_\mathrm{i}$) and $A_\mathrm{i}$ angular ones ($\{\mathfrak{A}(i)\}$, $\mu = 1, 2, ..., A_\mathrm{i}$). The radial functions describe a pair-wise interaction, whereas angular do so for three particles inside $R_\mathrm{c}$. Below the equations for these functions are written explicitly. First, a special cutoff function has the form

\begin{equation}
f_\mathrm{c}(R_{ij}) = 
\begin{cases}
\frac{1}{2} \left[ \cos{\frac{\pi R_{ij}}{R_\mathrm{c}} + 1} \right], & R_{ij} \leq R_\mathrm{c} \\
0, & R_{ij} > R_\mathrm{c}.
\end{cases}
\end{equation}

Now, there are three types of radial symmetry functions

\begin{equation}\label{eq.G1}
\mathfrak{G}_1(i) = \sum_{j \in \mathfrak{N}(i)} f_\mathrm{c}(R_{ij}),
\end{equation}

\begin{equation}\label{eq.G2}
\mathfrak{G}_2^{\mu}(i) = \sum_{j \in \mathfrak{N}(i)} e^{-\eta_\mu(R_{ij} - R_\mathrm{s})} \cdot f_\mathrm{c}(R_{ij}),
\end{equation}

\begin{equation}\label{eq.G3}
\mathfrak{G}_3^\mu(i) = \sum_{j \in \mathfrak{N}(i)} \cos({\kappa_\mu R_{ij}}) f_\mathrm{c}(R_{ij}).
\end{equation}

The use of the above radial functions with different parameters $\eta_\mu$ and $\kappa_\mu$ forms a set of $R_\mathrm{i}$ radial functions describing atom $i$. To our knowledge, all ready-to-use trained NN potentials in \texttt{MAISE} are designed for $R_\mathrm{i} = 8$ symmetry functions of the second type for each atomic type in a system. These functions do only differ by the value of $\eta_\mu$ parameter, the value of $R_\mathrm{s}$ being one and the same. The same is true for the cutoff radius $R_\mathrm{c}$.

The two types of angular symmetry functions read

\begin{equation}\label{eq.G4}
\mathfrak{G}_4^{\mu}(i) = 2^{1 - \zeta_{\mu_1}} \sum_{j,k \neq i} (1 + \lambda_{\mu_2} \cos \theta_{ijk} )^{\zeta_{\mu_1}} \cdot e^{-\eta_{\mu_3}(R_{ij}^2 + R_{ik}^2 + R_{jk}^2)} \cdot f_\mathrm{c}(R_{ij}) \cdot f_\mathrm{c}(R_{ik}) \cdot f_\mathrm{c}(R_{jk})
\end{equation}

and

\begin{equation}\label{eq.G5}
\mathfrak{G}_5^{\mu}(i) = 2^{1 - \zeta_{\mu_1}} \sum_{j,k \neq i} (1 + \lambda_{\mu_2} \cos \theta_{ijk} )^{\zeta_{\mu_1}} \cdot e^{-\eta_{\mu_3}(R_{ij}^2 + R_{ik}^2)} \cdot f_\mathrm{c}(R_{ij}) \cdot f_\mathrm{c}(R_{ik}),
\end{equation}
where $\mu$ stands for a multiindex $(\mu_1, \mu_2, \mu_3)$, $\cos \theta_{ijk}$ is the angle formed by atoms $i$, $j$ and $k$ with atom $i$ in the center of the angle, $\lambda$ parameters may only take two possible values $+1$ and $-1$. It seems like all the NN potentials in \texttt{MAISE} models use only the symmetry functions of type $4$, each input layer of an atomic NN contains $A_i = 43$ scalar values of $G_4^\mu(i)$ per atomic type. More detailed information on the BPNN models used in \texttt{MAISE} may be found on the official code's website\footnote{https://maise.binghamton.edu/wiki/nnet.html}.

What makes our BPNN implementation different from the one in \texttt{MAISE} is the use of automatic differentiation \cite{baydin2018automatic} for NN inference of atomic potential energies, forces and virials. We used utilities available in Eigen Tensor\footnote{https://libeigen.gitlab.io/eigen/docs-nightly/unsupported/eigen{\_}tensors.html} and Eigen Autodiff\footnote{https://libeigen.gitlab.io/eigen/docs-nightly/unsupported/group{\_}AutoDiff{\_}Module.html} libraries. For better understanding of our framework we provide below the expression for the force acting on atom $k$:

\begin{multline}\label{eq.force}
F_{k,\alpha} = - \frac{\partial E}{\partial R_{k,\alpha}} = - \sum_{i = 1}^N \frac{\partial E_i}{\partial R_{k,\alpha}} = -\sum_{i = 1}^N \sum_{\mu = 1}^{G_i} \frac{\partial E_i}{\partial \mathfrak{G}^\mu (i)} \cdot \frac{\partial \mathfrak{G}^\mu (i)}{\partial R_{k,\alpha}} \\ = -\sum_{i = 1}^N \sum_{\mu = 1}^{R_i} \frac{\partial E_i}{\partial \mathfrak{G}_2^\mu (i)} \cdot \frac{\partial \mathfrak{G}_2^\mu (i)}{\partial R_{k,\alpha}} -\sum_{i = 1}^N \sum_{\mu = 1}^{A_i} \frac{\partial E_i}{\partial \mathfrak{G}_4^\mu (i)} \cdot \frac{\partial \mathfrak{G}_4^\mu (i)}{\partial R_{k,\alpha}}.
\end{multline}

\begin{figure}[ht]
\centering
\includegraphics[width=0.7\textwidth]{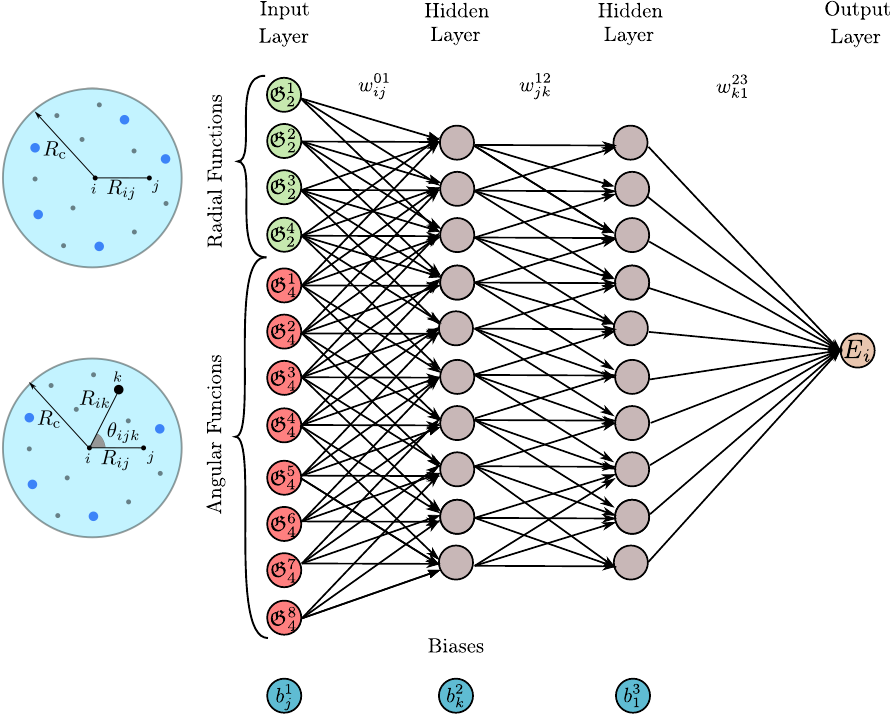}
\caption{Atomic fully connected Behler-Parrinello neural network architecture used in \texttt{MAISE}. On this schematic illustration the input layer consists of $R_i = 4$ scalar values as obtained from $2$nd type radial symmetry describing some atom $i$ plus $A_i = 8$ values for $4$th type angular functions. Then follows two intermediate hidden layers (here with 10 neurons each). The final quantity being predicted by this NN during the forward pass is atomic potential energy $E_i$. The backward pass evaluates gradients of $E_i$ with respect to the input layer's symmetry functions.}
\label{fig.AtomicNN}
\end{figure}

The final expression for the force $F_{k,\alpha}$ consists of two components: the first one is fully determined by the radial symmetry functions and the second one is by angular functions. The derivatives of symmetry functions with respect to atomic positions $\frac{\partial \mathfrak{G}_t^\mu (i)}{\partial R_{k,\alpha}} (t = 2, 4)$ are evaluated analytically and the derivatives of atomic energies with respect to symmetry functions $\frac{\partial E_i}{\partial \mathfrak{G}_t^\mu (i)} (t = 2, 4)$ are found via automatic differentiation. Automatic differentiation technique allows one to both infer the scalar value of atomic potential energy $E_i$ and immediately obtain input symmetry function layer gradients, see Figure \ref{fig.AtomicNN}. 

The analytical form of the atomic potential energy being evaluated in this work may be written explicitly:

\begin{equation}\label{eq.AtomicNN}
E_i = \mathfrak{F}_1 \left( b_1^3 + \sum_{k = 1}^{10} w_{k1}^{23} \cdot \mathfrak{F}_2 \left( b_k^2 + \sum_{j = 1}^{10} w_{jk}^{12} \cdot \mathfrak{F}_3 \left( b_j^1 + \sum_{\mu = 1}^{G_i} \mathfrak{G}^\mu(i) \cdot w_{ij}^{01} \right)  \right) \right),
\end{equation}
where $\mathfrak{F}_1$ stands for the activation function between the input layer and the first hidden layer, $\mathfrak{F}_2$ is so between the first and the second hidden layers and $\mathfrak{F}_3$ is the output activation layer. In \texttt{MAISE} $\mathfrak{F}_1(x) = \mathfrak{F}_2(x) = \tanh(x)$ and $\mathfrak{F}_3(x) = x$. The weights of the atomic NN are $w_{ij}^{01}$ between the input and the first hidden layer, $w_{jk}^{12}$ are between the first an the second hidden layers and $w_{k1}^{23}$ are between the second hidden and the final layer (with only one value). The corresponding model biases are given by $b_j^1$, $b_k^2$ and $b_1^3$.

\subsection{Software implementations}

To implement the ML-potentials described above, we added the following files to the folder with all the potentials in the project:

\begin{itemize}
    \item \texttt{DeePMD.h} and \texttt{DeePMD.cxx} --- \texttt{DeePMD} potentials support.
    \item \texttt{mlip4.h} and \texttt{mlip4.cxx} --- \texttt{MLIP4} potentials support.
    \item \texttt{maise{\_}bpnn.h} and \texttt{maise{\_}bpnn.cxx} --- \texttt{MAISE} potentials support.
\end{itemize}

For the latter potential there is a support for the \texttt{MAISE} model file parser along with the automatic differentiation utilities based on \texttt{Eigen} library. Both can be found in the corresponding subfolders in the \texttt{tools} folder.

\section{Applications}
\label{sec:examples}
Molecular dynamics with conventional interatomic potentials represents the standard use case and serves as a benchmark for any simulation package. Our implementation currently supports the LJ and EAM potentials. These models cover a wide spectrum of applications, from simple liquids to realistic metallic systems:
\begin{itemize}
	\item The MW stationary SW in argon step-by-step description 
	\item Conventional LJ and EAM potentials are employed for calculations of shock Hugoniot curves in argon and aluminum
	\item Comparative simulation using  machine learning potentials and EAM in copper
\end{itemize}    

\subsection{\label{sec:MW}Moving window simulation}

\begin{figure}[htbp]
	\centering
	\includegraphics[width=0.7\textwidth]{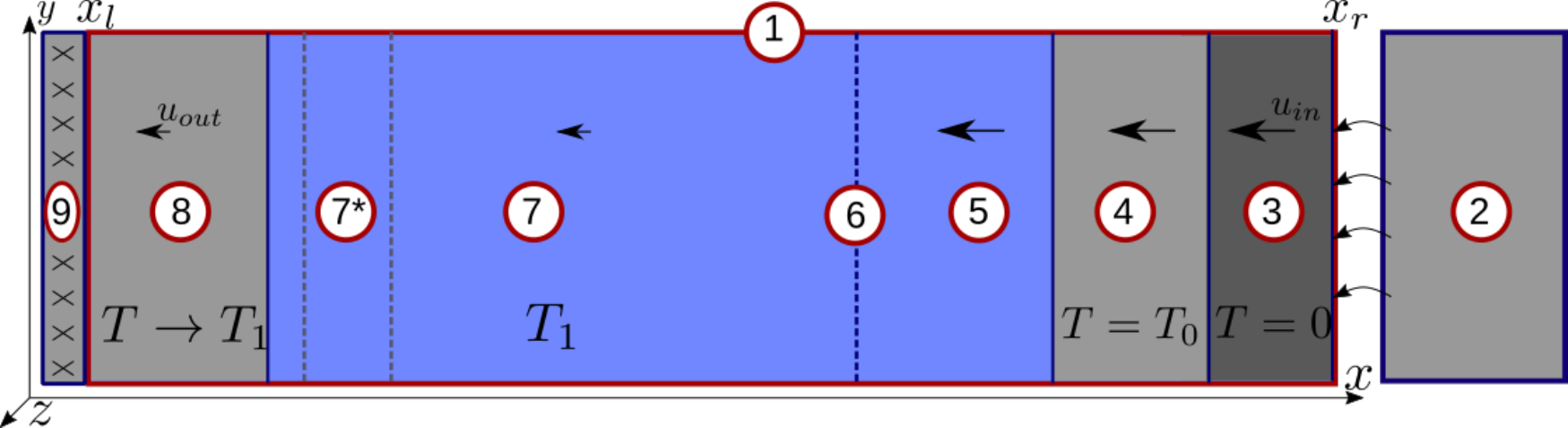}
	\caption{1 --- the moving window, 2 --- the basic sample prepared for inflow in the simulation box, 3 --- atoms with $T=0$, 4 --- the inflow thermostat region with parameters $(T=T_0, v=u_{in})$, 5 --- simulation domain of SW flow, 6 --- SW front; 7 --- relaxation behind the front, 7* --- statistics accumulation for thermostat target temperature update, 8 --- the outflow thermostat $(T\rightarrow T_1, v=u_{out})$, 9 --- atoms deleting.}
	\label{fig:scheme_mw}
\end{figure}
Many physical phenomena of interest, such as SWs, involve a localized disturbance propagating through the material. Simulating such systems with a conventional static simulation domain is highly inefficient, because computational cost grows as the square of the material sample length. 
 
The most of the computational effort is wasted on undisturbed parts of the material. To overcome this limitation, the MW technique is proposed, in which the simulation domain follows the propagating process. This technique has been studied accurately for both MD and SPH~\cite{Zhakhovskii:1997,Zhakhovsky:1999,Zhakhovsky:2011,Murzov:2021,MURZOV:2024}. The basic idea is that the velocity of the simulation domain is adjusted by the iterative feedback algorithm in order to establish the desired position of the shock front. Here, the algorithm is mostly inspired by the adaptive MW (AMW)~\cite{MURZOV:2024} developed for contact SPH and refined here for MD within a simplified algorithm of SW velocity $u_s$ determination. The iterative MW algorithm successively estimates the velocity of the SW and applies velocity transformations of the reference system to stop the movement of the wave inside the MW. The similarity of contact SPH and MD, in case of LJ and EAM interatomic potentials, is the locality of interactions between the atoms/particles, which allows similar implementations of the technique in the code. 
The benchmark for stationary SW MW simulation in argon is included in \emph{problems} providing step-by-step introduction to the stationary SW simulation. 

The necessary condition of a simplified AMW implementation is the definition of correct inflow/outflow boundary conditions. Additional Langevin thermostats applied at the inflow and outflow regions help to control velocities and suppress disturbances from the free surfaces. The initial simulation domain has dimensions $L_x\times L_y\times L_z$ in 3D. The periodical boundary conditions are imposed along $y$- and $z$-axes. Along $x$-axis there are inflow and outflow boundary conditions. The box is divided by several subdomains as shown in Fig.~\ref{fig:scheme_mw}. The simulation domain 1 is filled by replicating a basic sample 2 which is prepared in advance. To do this, we use periodic boundary conditions for all directions and relax the basic sample using Langevin thermostat at a temperature $T=T_0$ and average velocity $\mathbf{v}=0$; after that some NVE simulation is performed for tens of picoseconds in order ``to forget'' the thermostat.

To prepare the basic sample for modeling, a cubic simulation domain with approximate dimensions $8\times 8\times 8$~nm$^3$ is filled. The atoms are initially arranged in a face-centered cubic lattice. The lattice parameter is set to $\approx {0.462}\un{nm}$. The initial temperature of argon is $T_0=87\un{K}$, and the Lennard-Jones potential with parameters $\epsilon \approx 1.0312\un{kJ/mol}$ and $\sigma \approx 0.3384\un{nm}$ is used, with a cutoff radius of $0.8125\un{nm}$. The calculation with the basic sample is performed using a Langevin thermostat with a stiffness parameter $\beta=0.5\un{ps}$ for a duration of 12 ps, employing a time step of 4~fs to achieve thermal equilibrium before proceeding with further simulations related to SW. The atomic state of the basic sample is saved into a separate file (usually of HDF format~\cite{hdf}) that is used in the following routines.

At the inflow boundary, new atoms are inserted and move through the boundary, which is placed initially on the rightmost side of the simulated sample with $x=x_r$. It should be noted, that subsequently the inflow boundary position is varied to give room for the new layers of atoms generated after several integration time steps ($\approx$ 10), when the neighboring atoms lists are rebuilt. These new atoms are taken from the pre-equilibrated basic sample 2, moreover, the feeding of the basic sample is repeated periodically, when reaching the right boundary of the sample 2. The inflow subdomains 3 and 4 use Langevin thermostat with the same average velocity parameter $\mathbf{v}=(u_{in},0,0)$, but different target temperatures $T=0$ and $T=T_0$ correspondingly.

At the outflow boundary, atoms leaving the box are removed. In a stationary state, the number of inserted and deleted atoms is balanced, so that the total number of simulated atoms remains approximately constant over time. 
The outflow boundary has a fixed coordinate $x = x_l$. Joined to the MW subdomain 8 uses thermostat to force atoms to the prescribed outflow velocity $\mathbf{v}=(u_{out},0,0)$. At the same time the temperature of the material in this subdomain could be estimated \emph{a posteriori} and tends to the temperature in the SW $T_1$, this temperature  is calculated from the velocities of atoms in region 7* of subdomain 7 behind the SW front. The flow parameters are inferred using the statistical relation introduced in Appendix, Moreover, for a stationary SW profile, the certain characteristic $f(X)$ corresponds to the dependence of this quantity on time in some Lagrangian or fixed mass element as $f(t) = f(X/u_s)$, reflecting the self-similarity of the solution for the SW moving at a constant speed $u_s$.

We proceed to the algorithm for determining the shock wave speed.  
To adjust the velocity of the moving system, let us define the spatially averaged pressure that describes the position of the SW front within the simulation domain $F = P/P_t$.
Here $P$ is the current instant average pressure, determined in accordance with Appendix, and $P_t$ is the target average pressure corresponding to the desired approximately estimated position of the SW front. Thus, $F \approx 1$ is the target value. The value of $F$ fluctuates during calculations, which is accounted in the algorithm for adjusting the velocity $u_{out}$. The algorithm employs the average slope $\left<\frac{dF}{dt}\right>$, obtained through the linear fit using the least squares method over a set of function evaluations $F(t_m)$ over a sufficient time interval at moments $t_m$. In a steady-state, the slope $\left<\frac{dF}{dt}\right>$ approaches zero. The choice of value $F$ may be different; however, the convergence criterion for the algorithm to find the velocity of the reference system is the monotonic dependence of the slope $\left<\frac{dF}{dt}\right>$ on the velocity $u_{out}$ near the optimal value.  
The speed of such a reference system coincides with the SW speed $u_s$, so the inflow velocity of atoms is $u_{in} = -u_s$. The outflow velocity of atoms $u_{out}$ is related to $u_{in}$ by the relation $u_{out} - u_{in} = u_p$ and it is sufficient to determine the velocity $u_{out}$.  

\begin{figure*}[htbp]
	\centering
	\includegraphics[width=\textwidth]{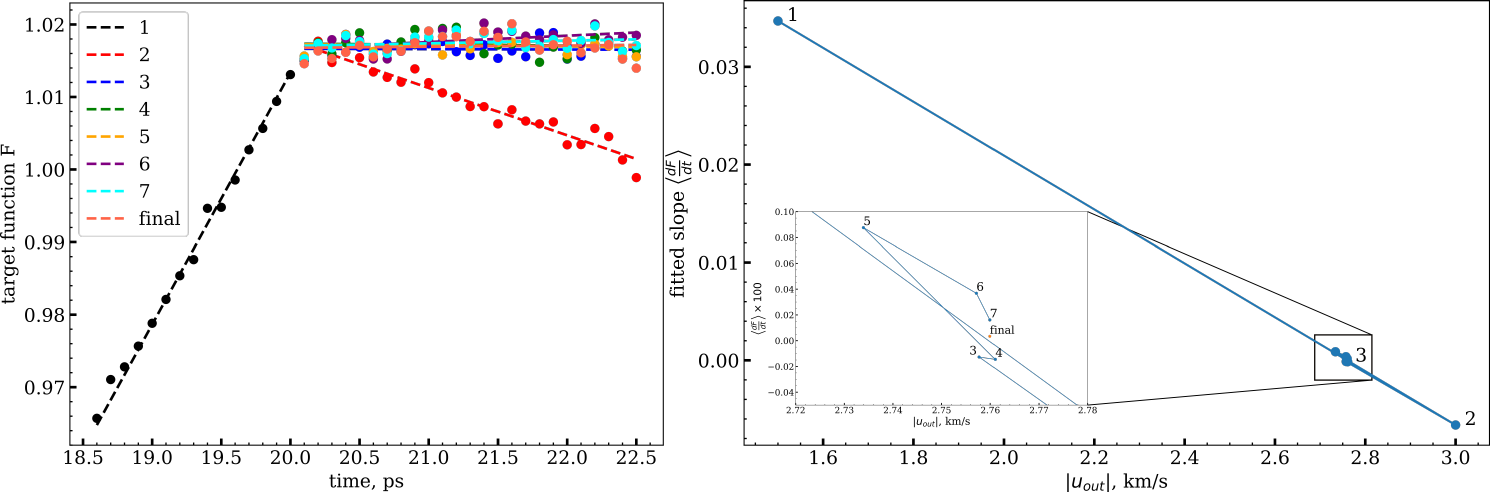}
	\caption{Left: The convergence of MW algorithm for stationary flow in SW simulation is shown for argon with starting outflow velocity $u_{out}=1.5$~km/s. The measured values  target function $F$ with correspondent fit    are enumerated for iterations of the MW dichotomy search. Right: Linear fit slopes of the target function in the sequence of restarting simulations converges to zero.}
	\label{fig:convergence}
\end{figure*}
The initialization of the algorithm involves simulating a SW propagating with a specific fixed speed behind the front in such a system that the wave moves from the outflow boundary to the inflow boundary. However, the outflow velocity must be less than zero $u_{out}<0$ to ensure that by the time the shock wave reaches the desired position ($F = 1$), the exit region already corresponds to the state of shock compression---pressure, density, and internal energy should satisfy the Hugoniot relations. In this simulation the rate of increasing the target function is maximal usually as shown in Figure~\ref{fig:convergence} for the iteration 1 with black points. 

Once such a sample is obtained, we save the current state to a separate HDF-file and the algorithm for estimating the front velocity is initiated. The algorithm for finding the reference system is a master script that runs simulations by sequentially comparing the latest pair of slope values of the function $\left<\frac{dF}{dt}\right>$ for two different velocities of the system, determined, for example, as $u_{out}$. 
The algorithm consists of the following steps:  
\begin{enumerate}  
	\item Define the boundaries for outflow velocity $u_{out}$: $u_{out}^{min}$ and $u_{out}^{max}$.  
	\item Using separate simulation restarts get the signs of the average slope $\left<\frac{dF}{dt}\right>$ at $u_{out}^{min}$ and $u_{out}^{max}$. In the optimal case $\left<\frac{dF}{dt}\right> = 0$, while the search for velocity $u_{out}$ is performed using the dichotomy. Simulations are run at different values of $u_{out}$ starting from the saved state obtained upon reaching the target value $F = 1$. The search for the root of the equation $\left<\frac{dF}{dt}\right>(u_{out}) = 0$ using the dichotomy method terminates when the search range $|u_{min}-u_{max}|$ narrows to less than $1$~m/s.
	\item Since the obtained state may still differ from the stationary one, because there are still relaxation, the value of $u_{out}$ is repeated in step 2 with double the simulation time. The resulting state is then saved.  
	\item Finally, the dichotomy algorithm is restarted to determine the velocity $u_{out}$ more precisely.
\end{enumerate}  
In the example of this script for a shock wave SW the parameters are $u_p=-2\un{km/s}$ and the search begins at $u_{out}=-1.5\un{km/s}$ to observe the SW moving from right to left.
The convergence of algorithm for the discussed MW simulations is shown in Figure~\ref{fig:convergence}.

The proposed speed adjustment algorithm differs from that described in~\cite{MURZOV:2024}, enabling profiles to be obtained in less simulation time and requiring fewer initial parameter adjustments. The dichotomy method relies solely on the monotonicity of $u_s(u_p)$ and the necessity to choose an initial velocity search interval $[u_{out}^{min}, u_{out}^{max}]$ such that the measured values of $\left<\frac{dF}{dt}\right>$ at the interval's endpoints which have opposite signs. The second constraint is to estimate $\left<\frac{dF}{dt}\right>$ over a sufficiently long time period $L/u_{in}$, where $L$ represents several sizes of the base sample being fed, to exclude the influence of oscillations caused by structural peculiarities of the inflow sample and the structures of the SW front of chosen amplitude on measurements of the target function $F$.

\subsection{Argon shock Hugoniot: system of isotherms and a series MW simulations}  	
\begin{figure}[htbp]
  \centering
  \includegraphics[width=0.7\textwidth]{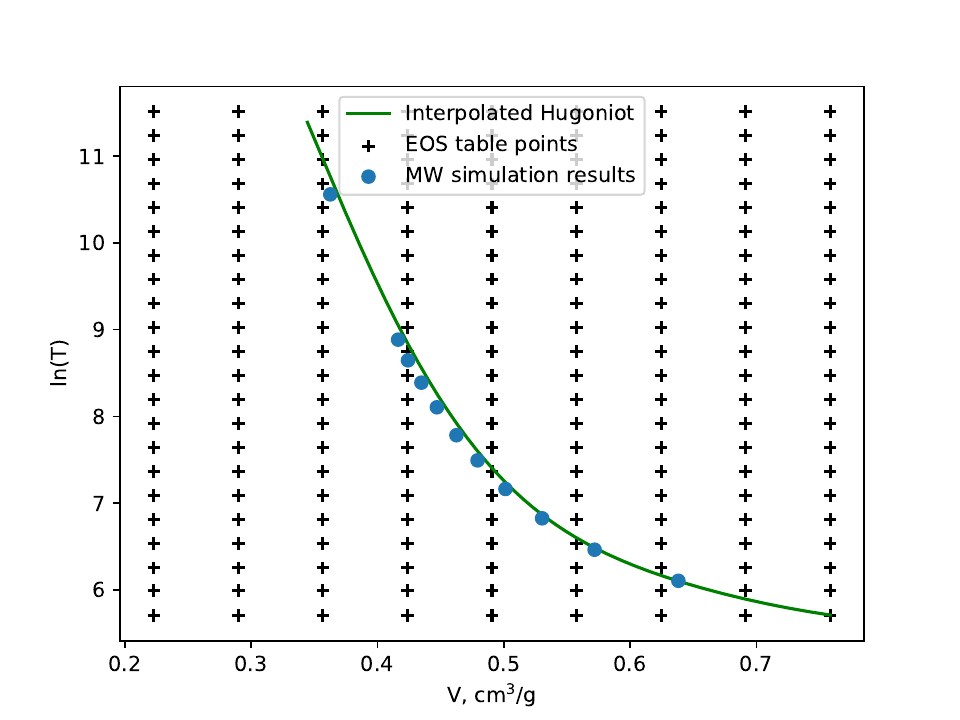}
  \caption{Shock Hugoniots for the Lennard--Jones system obtained with fixed-box and moving window simulations.}
  \label{fig:lj_hugoniot}
\end{figure}

In the case of Lennard--Jones, the potential parameters $\sigma$ and $\epsilon_{0}$ can be directly set in the Python script.
As a first demonstration, we obtain an EOS table for the Lennard--Jones system by simulating cubic samples at different temperatures and volumes (fixed box simulations). The cubic sample contains 6750 atoms of mass $40\un{g/mol}$. The initial state of the Hugoniot is the point $T_0=300\un{K}$ and  $V_0=340\un{nm^3}$. We perform a series of simulations to reach a state with given specific volume-temperature state $(V_i, T_i)$ for $V_i \in [100, 340]\un{nm^3}$, while $T_i$ set in the following range $[300, 10^{5}]\un{K}$. The cubic sample dimensions are varied to achieve given volumes $V_i$ and define the specific volume for a fixed atoms number. Each simulation runs for 60~ps using Langevin thermostat with the stiffness parameter $\beta=0.5\un{ps^{-1}}$. The time step is a constant for each simulation run, but inversely depends on the target temperature of thermostat $\Delta t = 10^{-2}\sqrt{10/T_i}\un{fs}$.
Such a choice saves CPU time for low--temperature simulations, while providing accuracy to the high-velocity atoms as temperature grows. The selected time--step formula gives the standard deviation of kinetic energy of less than 2\% in NVE ensemble simulation. After 20~ps equilibration, instant values of pressure, energy, temperature and density (see Appendix) collected each 100 steps are averaged to diminish fluctuations. The shock Hugoniot in Fig.~\ref{fig:lj_hugoniot} is calculated as a locus of points where the left--hand side of the Rankine--Hugoniot equation \ref{eq.RH} $H=0$. Bilinear interpolation is used to solve Eq.~\ref{eq.RH} using the calculated values $H_i(v_i,\epsilon_i,P_i)$, where $v$ and $\epsilon$ are the specific volume and energy respectively. 
For each grid point $(v_i, T_i)$ the function $H$ is evaluated, using the values $(v_i,\epsilon_i,P_i)$ as well as the initial point data $(v_0,\epsilon_0,P_0)$: 
\begin{equation}\label{eq.RH}
H(v,\epsilon,P) = \epsilon - \epsilon_0 - \frac{P+P_0}{2}(v_0-v).
\end{equation}

\begin{figure}[htbp]
  \centering
  \includegraphics[width=0.7\textwidth]{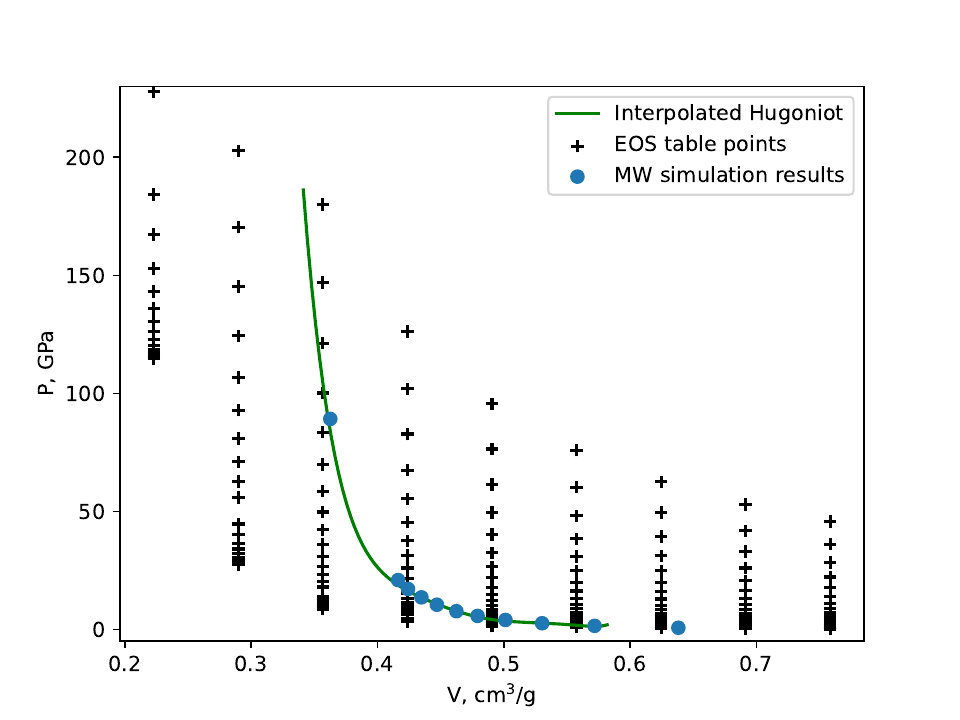}
  \caption{Shock Hugoniot for the Lennard--Jones system obtained with fixed-box and moving window simulations.}
  \label{fig:lj_hugoniot_p}
\end{figure}

The MW simulations (see Section~\ref{sec:MW}) are run in the simulation box with dimensions $112\times 8\times 8\un{nm^3}$ and the initial state $(v_0,\epsilon_0,P_0)$ previously selected for the Rankine--Hugoniot equation. In these simulations we use the following set of particle velocities behind the shock front: $u_p =  0.25$, 0.5, 0.75, 1.0, 1.25, 1.5, 1.8, 2.1, 2.4, 2.7~km/s. The MW simulation results shown with blue circles in Figures \ref{fig:lj_hugoniot} and \ref{fig:lj_hugoniot_p} are consistent with the results of Eq.~\ref{eq.RH} numerical solution.

\subsection{Comparison of aluminum shock Hugoniot in the MW using different EAM potentials}  
\begin{figure}[htbp]
  \centering
  \includegraphics[width=0.7\textwidth]{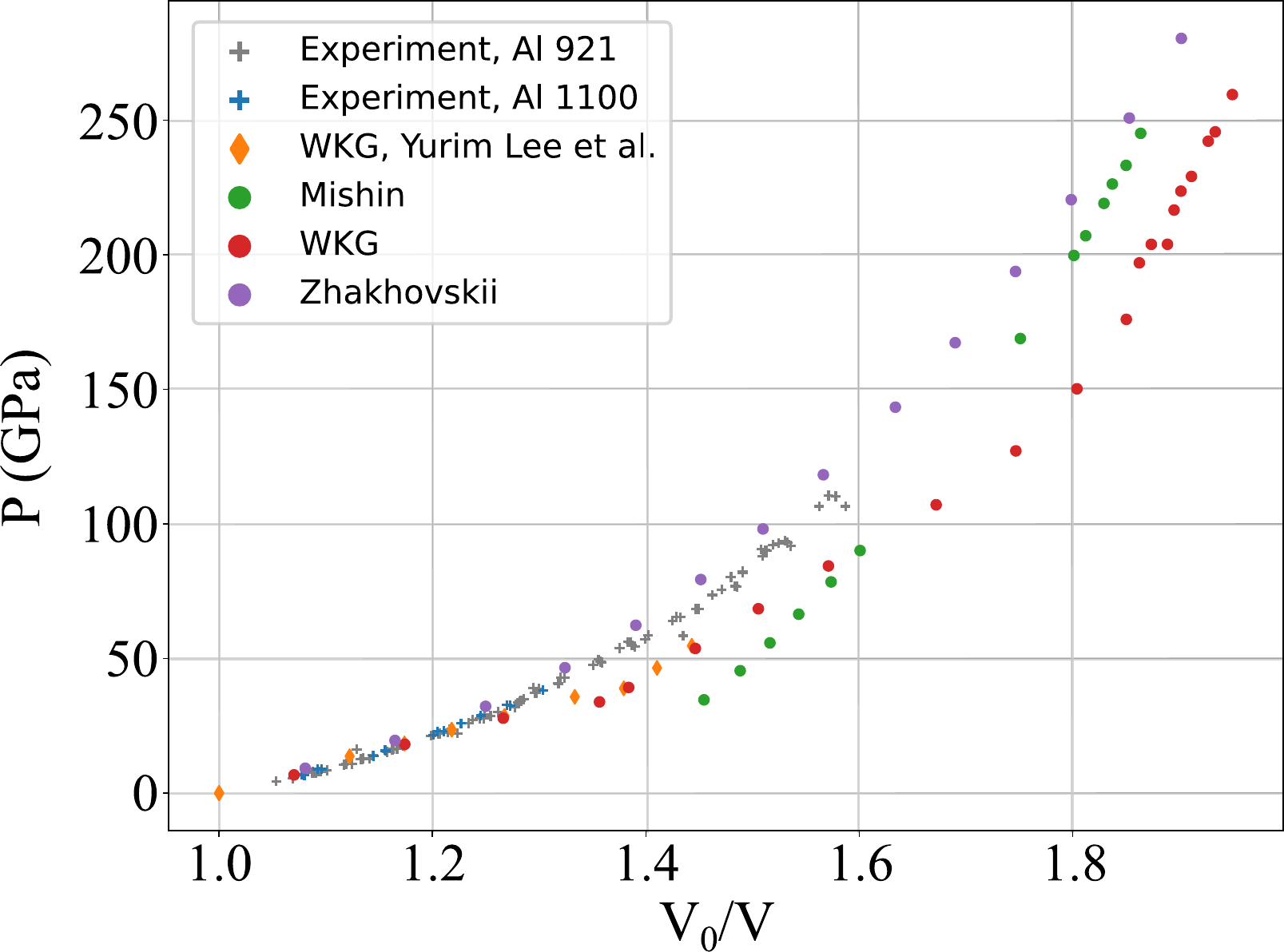}
  \caption{Shock Hugoniots of aluminum simulated with different EAM potentials using the moving window method. Crosses represent experimental data \cite{Marsh1980LaslSH}. Diamond -- simulation data from Pu et al. (2020) \cite{WKG}. The three potentials -- Mishin \cite{MISHIN20041451}\cite{Becker2011}, Zhakhovskii \cite{ZHAKHOVSKII20099592} and WKG \cite{Winey_2010} -- were taken from the Interatomic Potentials Repository. }
  \label{fig:al_eam}
\end{figure}

The EAM potential is loaded from a tabulated file, with support for single-element metals. This allows for modeling of crystalline solids and capturing many-body interactions beyond the reach of simple pair potentials. For instance, in aluminum under shock loading, the moving window approach enables efficient simulation of the Hugoniot curve and structural evolution while keeping the particle number constant. The results of MW simulations with different EAM potentials are presented in Fig \ref{fig:al_eam}. The potentials were taken from the Interatomic Potentials Repository. \footnote{https://www.ctcms.nist.gov/potentials/entry/2004--Mishin-Y--Ni-Al/},\footnote{https://www.ctcms.nist.gov/potentials/entry/2009--Zhakhovskii-V-V-Inogamov-N-A-Petrov-Y-V-et-al--Al/},\footnote{https://www.ctcms.nist.gov/potentials/entry/2009--Winey-J-M-Kubota-A-Gupta-Y-M--Al/}

\begin{table}[h]
	\centering\caption{Parameters for MW simulations in aluminum}
	\begin{tabular}{ccccc}
		\hline
		Potential & lattice parameter $a$, nm & range of $u_p$, km/s \\
		\hline
		Zhakhovsky & 0.4032 & [0.5, 7] \\
		\hline
		Mishin & 0.4024845356 & [2.1, 6.86] \\
		\hline
		WKG & 0.4055 & [0.5, 6.558] \\
		\hline
	\end{tabular}\label{table1}
\end{table}
The simulation domains were prepared independently for each potential because of slightly different initial density of the material under normal pressure and temperature. The inflow basic sample dimensions were approximately $L_x\times L_y\times L_z=8\times 8\times 8\un{nm^3}$. The sample consisted of the atoms arranged in a face-centered cubic lattice and was relaxed during 20~ps using Langevin thermostat with a stiffness $\beta=0.5\un{ps^{-1}}$. The moving window box was the basic sample replicated 50 times. Three series of simulations were performed for different velocities of SW piston shown in Table~\ref{table1}. The starting outflow velocity was taken -1.5~km/s being less than the sound speed in the bulk aluminum. 

Together, the LJ and EAM examples highlight the ability of the package to handle both minimalistic and realistic interatomic models through the same Python interface, while the optimized C++ backend ensures computational efficiency.

\subsection{Molecular dynamics with machine learning potentials}

\begin{figure}[htbp]
  \centering
  \includegraphics[width=1.0\textwidth]{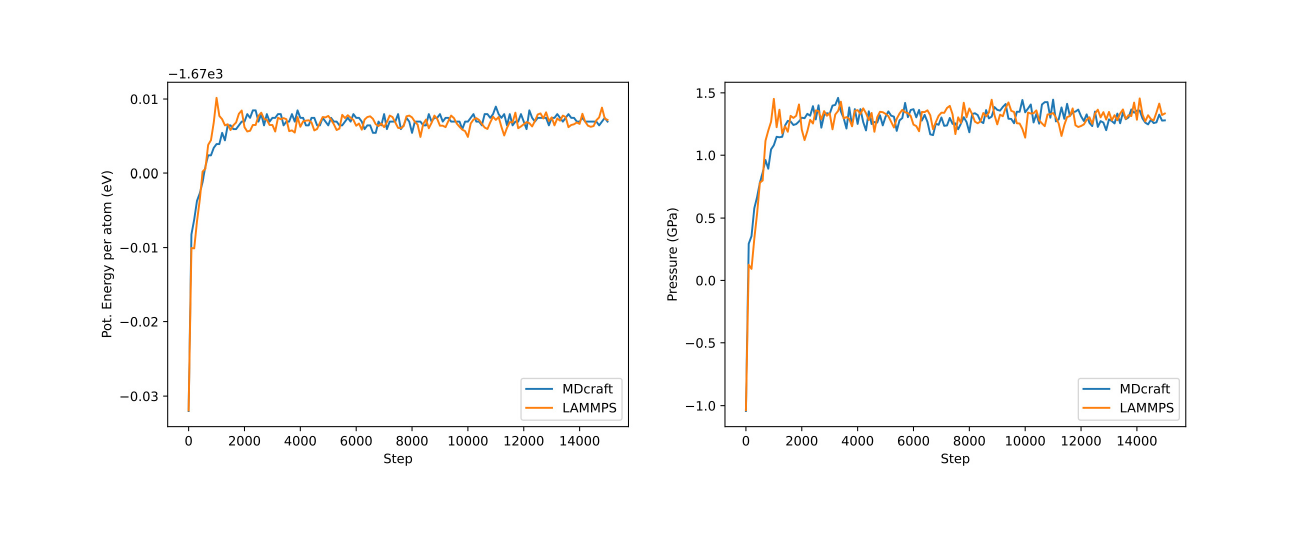}
  \caption{Comparison of thermodynamical properties along some MD-trajectory in \texttt{DeePMD} and \MDcraft codes.}
  \label{fig:deepmd_cmp}
\end{figure}

To demonstrate the correctness of the currently supported ML-potentials we show in Fig.~\ref{fig:deepmd_cmp} the comparison between potential energy and pressure along some MD-trajectory both for \MDcraft and \texttt{LAMMPS} implementations. The example is given for the copper \texttt{DeePMD} potential which can be freely downloaded from the Interatomic Potentials Repository\footnote{https://www.ctcms.nist.gov/potentials/system/Cu/}. Such simulations are commonly used to validate MD algorithms: energy conservation in the microcanonical ensemble, correct thermostatting in the canonical ensemble, and the ability to reproduce well-known structural observables.
The chosen potential corresponds to the one introduced in Ref.~\cite{Cu-model-2023}. It is clearly seen that both codes provide the same system evolution: the initial out-of-equilibrium regions and the switch to the thermodynamic equilibrium regime are nearly the same. The number of atoms in this test was taken to a rather small number of $2048$ just to compare the output of the two codes. Both calculations were launched in NVT ensemble with $T=300$~K and copper density of $\rho=8.82$~g/cm$^3$. The timestep was taken equal to $4$~fs which is a rather reasonable value for most MD calculations in solid metallic systems. The main difference in these two simulations is attributed to distinct initial atomic velocity distributions that are set randomly according to the gaussian distribution with a given temperature $T$.

\begin{figure}[htbp]
  \centering
  \includegraphics[width=0.7\textwidth]{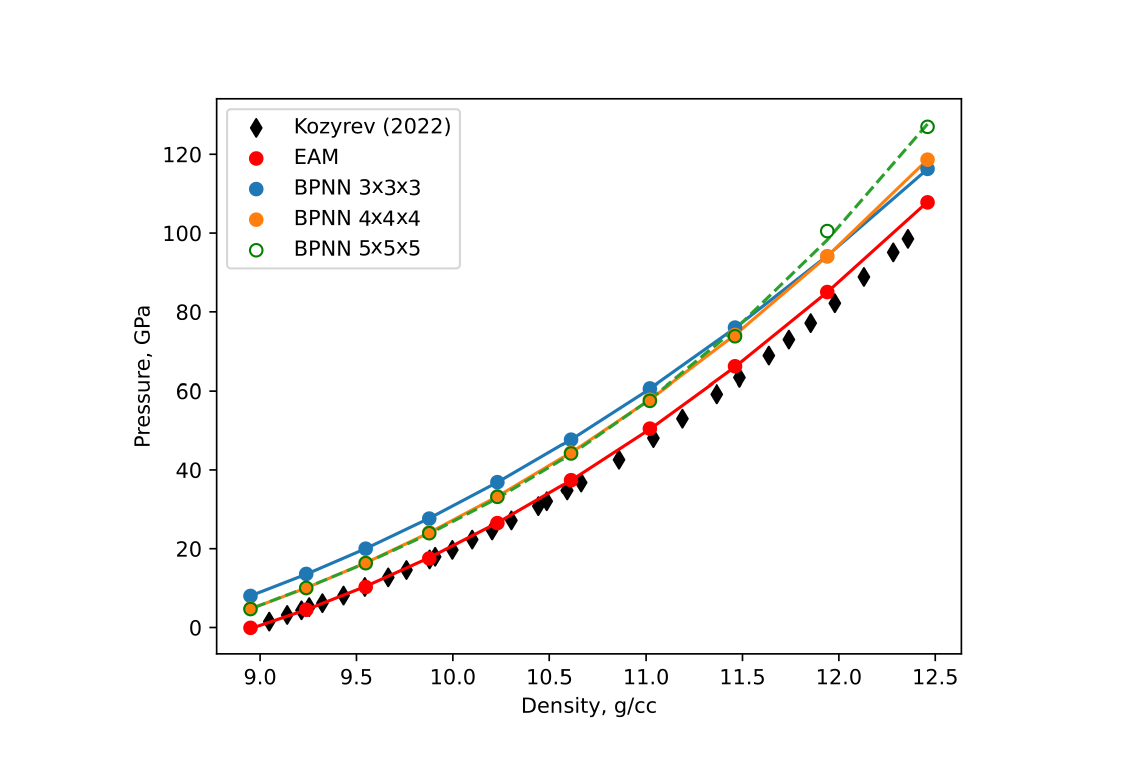}
  \caption{The cold compression curve for Cu calculated with BPNN potential in \MDcraft, compared with the experimental results~\cite{kozyrev2023thermodynamic} and our EAM potential~\cite{Perriot-2014} calculations. The points are actual MD simulation results and lines are smooth approximations.}
  \label{fig:cu_bpnn}
\end{figure}

Additionally one can see in Fig.~\ref{fig:cu_bpnn} the cold compression curve of copper as obtained from our support for MD with BPNN potentials (for different supercell sizes) and compared with the experimatntal data taken from Ref.~\cite{kozyrev2023thermodynamic} and our calulations with EAM potential as described in Ref.~\cite{Perriot-2014}. We used \texttt{nn{\_}Cu{\_}d3{\_}v0} BPNN potential as given in the \texttt{models} folder of \texttt{MAISE}. The overall agreement is observed between all the given results. The data points obtained with the mentioned BPNN potential in \MDcraft calculations are in perfect agreement with those from the native MD implementation of \texttt{MAISE}. A better agreement with the reference data is observed for the EAM potential (for a face-centered cubic supercell $24 \times 24 \times 24$ containing $55296$ atoms in total), and the overall correct trend is observed for the curves with the chosen particular  pretrained BPNN model for different supercell sizes between $3 \times 3 \times 3$ and $5 \times 5 \times 5$. Due to a more complicated architecture of the supported BPNN potentials—which are three-body potentials and notably computationally expensive—we limited out tests to relatively small numbers of atoms (from $108$ in the smallest supercell up to $500$ in the largest one). In contrast, the potential in the EAM form is substantionally more efficient.

\section*{Conclusions}
The article concludes by introducing \MDcraft as a versatile and scalable simulation software designed to address the complexities faced by users of traditional MD packages. It is structured using a modular approach with core modules tailored for specific tasks, enhancing flexibility and scalability. By leveraging Python scripting, \MDcraft seamlessly integrates machine learning tools, eliminating the need for low-level coding. While existing MD packages are robust, they often require extensive API learning, which can be daunting. In contrast, \MDcraft offers a high-level Python API, simplifying the incorporation of advanced techniques such as ML potentials and enabling efficient parallel simulations on modern clusters. This design allows users to handle complex tasks like moving window simulations without modifying core modules. The package is well-suited for meeting the demands of contemporary computational research, providing an accessible and powerful tool for MD simulations.

\section*{Acknowledgments}
The work was supported by the Russian Science Foundation grant 24-19-00746.
\section*{Appendix. Analysis of MD simulation results}

Flow and thermodynamic parameters can be obtained, according to their statistical definitions~\cite{Yip:2005}, from the analysis of positions $\mathbf{r}_i$, velocities $\mathbf{v}_i$ and forces $\mathbf{f}_i$ of atoms. 
Let's consider the system of $N$ atoms, for which the equations of motion Eq.~(\ref{eq:MD-system}) are solved.
The sample under study is divided into a set of sufficiently large volumes $V_l$ containing $N_l$ atoms, where statistical characteristics are calculated. The total mass of atoms in such a volume is obviously
$$
M_l = \sum_{i=1}^{N_l} m_i,
$$
and then the potential energy per unit mass is given by the additive expression:
$$
E_{l}^{\text{pot}} = \frac{\sum_{i=1}^{N_l} U_i}{M_l}.
$$

The temperature on each degree of freedom $\alpha = x, y, z$ of atoms is defined by the kinetic energy in the reference frame of the immobile center of mass of the selected volume $l$:
$$
E_l^{\text{kin}} = \sum_{i=1}^{N_l} m_i (u^\alpha_{i} - u^\alpha_l)^2 / 2.
$$

The internal energy of Lagrangian particles in hydrodynamics corresponds to the total energy, which is the sum of potential and kinetic energy in the center-of-mass system:
$$
E_l^{\text{int}} = (E_{l}^{\text{pot}} + E_l^{\text{kin}}) / M_l.
$$

The relationship between the kinetic energy $E_l^{\text{kin}}$ and temperature is established using the specific heat capacity per atom for each degree of freedom, which is the half of Boltzmann constant $k_b/2$:
$$
T^{\alpha}_l = \frac{2 E_l^{\text{kin}}}{k_b N_l},
$$
where the components $u^\alpha_l$ of the velocity vector $\mathbf{u}_l$ of the center of mass of the selected $N_l$ atoms are used:
$$
\mathbf{u}_l = \frac{1}{\sum_{i=1}^{N_l} m_i}
\begin{bmatrix}
\sum_{i=1}^{N_l} v^x_i m_i \\
\sum_{i=1}^{N_l} v^y_i m_i \\
\sum_{i=1}^{N_l} v^z_i m_i
\end{bmatrix}.
$$

For MW simulation, the temperature profile averages $\langle T^{\alpha}_l\rangle$ are obtained using additional averaging in the stationary frame. In that case, data can be averaged over time moments $t = 1, M$, where in each moment, there may be different numbers of atoms $N_l(t = m)$ in the  reference volume $V_l$:
$$
\frac{1}{k_b \sum_{m = t_1}^{t_M} N_l(t = m)} \sum_{m = t_1}^{t_M} \sum_{i=1}^{N_l(t = m)} m_i (u^\alpha_{i}(t = m) - u^\alpha_l(t = m))^2 \stackrel{\text{des}}{=} \frac{1}{k_b} \left< \sum_{i=1}^{N_l} m_i (u^\alpha_{i} - u^\alpha_l)^2 / N_l \right>.
$$

The temperature is the average of the calculated components $T = (T^x + T^y + T^z) / 3$.
The pressure tensor $P^{\alpha \beta}$ ($\alpha, \beta = x, y, z$) is calculated as follows:
$$
P_l^{\alpha \beta} = \frac{1}{V_l} \left[ \left< \sum_{i=1}^{N_l} m_i (u_i^{\alpha} - u_l^{\alpha})(u_i^{\beta} - u_l^{\beta}) \right> + \frac{1}{2} \left< \sum_{i=1}^{N_l} \sum_{j \neq i}^{N^i_{\text{neibs}}} f^{\beta}_i (r^{\alpha}_{i} - r^{\alpha}_{j}) \right> \right].
$$
There are two terms: the first is related to the thermal motion of atoms, while the second term is the virial sum $W^{\alpha\beta}_l$ of tensor multiplication of the forces $\mathbf{f}_i$ and coordinate vectors $\mathbf{r}_{ij}$:
$$
W_l^{\alpha \beta} = \frac{1}{2} \sum_{j \neq i}^{N^i_{\text{neibs}}} f^{\beta}_i (r^{\alpha}_{i} - r^{\alpha}_{j}).
$$
The hydrostatic pressure is determined through the trace of $P_l^{\alpha \beta}$, $P_l = (P^{xx} + P^{yy} + P^{zz}) / 3$. Then one can obtain the relationship between pressure, temperature and virial:
$$
P_l = \frac{1}{N_a} \left( N_l k_b T_l - \frac{W_l^{\alpha \alpha}}{3} \right) / V_l.
$$

Generally speaking, there is no pressure and temperature isotropy immediately behind the SW front, and the relaxation occurs only at a sufficient distance from the front, where thermalization is established across all degrees of freedom.

\end{document}